\documentclass[aps,prl,twocolumn,groupedaddress]{revtex4}
\usepackage{bm}
\usepackage{amsthm}
\usepackage{amsmath}
\usepackage{graphicx}
\usepackage{verbatim}

\newtheorem{definition}{Definition}

\newtheorem{conv}{Convention}

\newtheorem*{lemma}{Simple Fact}
\newtheorem{claim}{Claim}

\newcommand{\inner}[2]{\ensuremath{\left\langle{#1,#2}\right\rangle}}

\begin{document}

%%%%%%\title{Analyzing the correlations for spin-$\frac{1}{2}$ particles and singlet pairs} 
\title{Bell inequality violations under reasonable and under weak hypotheses}

\author{Edson de Faria} 
\affiliation{Instituto de Matem\'atica e Estat\'\i stica, USP, S\~ao Paulo, SP, Brazil}
\author{Charles Tresser} 
\affiliation{IBM, P.O.  Box 218, Yorktown Heights, NY 10598, U.S.A.}
\email[]{edson@ime.usp.br}
\email[]{charlestresser@yahoo.com}

\date{\today}

\begin{abstract}

Given a sequence of pairs $(p_i,\overline{p}_i)$ of 
spin-$\frac{1}{2}$ particles 
in the singlet state, assume that Alice measures the normalized projections $a_i$ of the
spins of the $p_i$'s along vector ${\bm a}$ while Bob measures the
normalized projections $b_i$ of the spins of the $\overline{p}_i$'s
along vector ${\bm b}$.  Then \emph{Quantum Mechanics} (\emph{QM}) lets one evaluate
the correlation $\inner{a}{b}$ as $-\cos (\theta _{\bm a}- \theta _{\bm b})$ where $\theta _{\bm v}$ is
the angle between the vector ${\bm v}$ and a reference vector chosen
once and for all, and we also assume that all vectors are chosen in a
fixed plane.  Assuming \emph{Classical Microscopic Realism} (\emph{CMR})
there exist also normalized projection pairs $(a'_i, b'_i)$ of the spins
of the pairs $(p_i,\overline{p}_i)$ along the vector pair $({\bm a'},
{\bm b'})$.  Assuming QM and CMR, we also have
$\inner{a'}{b'}=-\cos(\theta _{\bm a'}- \theta _{\bm b'})$.     
Since all projections are in $\{-1,1\}$, 
$|\inner{c}{d}+\inner{c}{e}| + |\inner{f}{d}-\inner{f}{e}|\leq 2$ for 
$c$, $d$, $e$ and $f$ elements of
$\{a,b,a',b'\}$.  Assuming Locality (the impossibility of any effect of an event on
another event when said events are spatially separated) beside QM and CMR, Bell's theory
lets one deduce various violations of this inequality at some choices of quadruplets  
$Q\equiv ({\bm a}, {\bm b},{\bm a'}, {\bm b'})$.  Our main result is the existence of quadruplets $Q$'s where at least one of  
the above inequalities is violated if one only assumes QM, CMR and some very mild further hypotheses.  
These weak hypotheses only concern the behavior of correlations that we use near special $Q$'s.
\end{abstract}

%\keywords{}
\maketitle

%%%%%%%%%%%%%%%%%%%%%%%%%
We consider sequences of \emph{EPRB pairs}, \emph{i.e.,}
pairs of spin-$\frac{1}{2}$ particles $(p,\overline{p})$ whose wave
function's spin part is the \emph{singlet state}:  
\begin{equation}\label{Singlet} 
\Psi=\frac{1}{\sqrt{2}}(| +\rangle _p\otimes|
-\rangle_{\overline{p}}-| -\rangle_p\otimes|+\rangle_{\overline{p}}
)\,. 
\end{equation}
The EPRB name evokes a reformulation by Bohm \cite{Bohm},
using spin-$\frac{1}{2}$ particles pairs, of the so-called \emph{``EPR paper"} \cite{EPR}.  
The singlet state is an example of
\emph{entanglement} because the sum of tensor products in
(\ref{Singlet}) cannot be rewritten as a single tensor product of
one-particle states. 

Consider a sequence of EPRB pairs $(p_i,\overline{p}_i)$ and assume that Alice measures the
normalized projections $a_i$ of the spins of the $p_i$'s along vector
${\bm a}$ while Bob measures the normalized projections $b_i$ of the
spins of the $\overline p_i$'s along vector ${\bm b}$. Then by
\emph{Quantum Mechanics} (\emph{QM}) the correlation
$\inner{a}{b}$ is given by the \emph{Twisted Malus Law}: 
\begin{equation}\label{Correl1}
\inner{a}{b}\equiv \lim_{n\to\infty}\frac{1}{n} \sum_{i=1}^n
a_i b_i=-\cos(\theta _{\bm a}- \theta _{\bm b}) 
\end{equation}
where for any vector ${\bm v}$, the angle $\theta _{\bm v}$ is
measured relative to a reference vector chosen once and for all in a  
plane that is also fixed once and for all. 

Instead of assuming Predictive Hidden Variables (PHVs) as Bell in
\cite{Bell} one tends now following Stapp \cite{Stapp1985} (see also
\cite{Stapp1971}) and \cite{Leggett2008}) to make weaker realist
assumptions that may go collectively under the name of
\emph{Classical Microscopic Realism} (or \emph{CMR}) - CMR tells us \emph{e.g.,} that all the observables that could be measured have well defined
values - to which we have adapted the following conventions whose
origin goes back at least to \cite{Bell}.

\begin{conv}  Whenever we assume that QM is augmented by a form of CMR,
  we implicitly postulate that any n-tuple of quantities that are not
  measured, but that exist according to said form of CMR, has the
  values that would have been obtained if this n-tuple of quantities
  had been the one measured, the World being otherwise unchanged. 
\end{conv}

\begin{conv}  Whenever we assume that Quantum Mechanics is augmented
  by a form of CMR, we assume that the augmentation by said form of CMR
  is made \emph{without changing the statistical predictions of QM}. 
\end{conv}

It follows, assuming QM and CMR, the correlation of the
sequence of normalized projections $a_i'$ of the spins of the $p_i$'s
along vector ${\bm a'}$ with the sequence of normalized projections
$b_i'$ of the spins of the $p_i$'s along vector ${\bm b'}$ can as well
be computed from the twisted Malus Law as: 
\begin{equation}\label{Correl2}
\inner{a'}{b'} \;=\; -{\bm a'}\cdot{\bm b'}\,=\,-\cos(\theta
_{\bm a'}- \theta _{\bm b'})\,. 
\end{equation}

Bell assumes \emph{Locality} in the derivation of the main result of
\cite{Bell}.  We like the following formulation, but, 
like for CMR,
%%%%here and for Realism, 
%%%expert 
readers %%%%will probably 
may %%%%%
substitute their own preferences. 
\begin{definition}[Locality]  Assume that the space-time locations
  $({\bm{x}}_0, t_0)$ and $({\bm{x}}_1, t_1)$ are spatially separated
  (i.e. $\Delta {\bm{x}}^2> c^2 \Delta t^2$, where $c$ stands for the
  speed of light). Then \emph{Locality} tells us that the output of a
  measurement made at $({\bm{x}}_0, t_0)$ (respectively $({\bm{x}}_1,
  t_1)$) (or a CMR dependent value)  cannot depend upon the setting of
  a measurement tool at $({\bm{x}}_1, t_1)$ (respectively
  $({\bm{x}}_0, t_0)$). 
\end{definition} 

Given any four sequences ${u_i}, {v_i}, {x_i}, {y_i}$ with values in
$\{-1,1\}$, under genericity assumptions to the effect that all needed
limits exist, one has the following inequalities (where the roles of
the sequences can be exchanged since the sequences are all of the same
nature):  
\begin{equation}\label{CHSH0}
|\inner{u}{v} +\inner{u}{x}| + |\inner{y}{v}-\inner{y}{x}|\leq 2\,. 
\end{equation}
The inequalities (\ref{CHSH0}) contain the Bell
inequality from \cite{Bell}: 
\begin{equation}\label{Bell0}
\inner{v}{x} + |\inner{y}{v} -\inner{y}{x}|\leq 1\ .
\end{equation}
(with $u=v$ in \eqref{CHSH0} one gets $|\inner{u}{v} + \inner{u}{x}|=\inner{v}{x} +1$, hence 
\eqref{Bell0}).  
Inequalities with three
correlations like (\ref{Bell0}) are not well adapted when dropping the
Locality assumptions for reasons that should soon be clear.    
 
The two forms of the \emph{CHSH inequalities} \cite{CHSH69} are
contained in  (\ref{CHSH0}):   
\begin{equation}\label{eq:CHSH}
|\inner{a}{b} \pm \inner{a}{b'}| + |\inner{a'}{b} \mp \inner{a'}{b'}|\leq 2\ , 
\end{equation}
where the sequences $a$, $b$, $a'$, $b'$, are
as defined above.  The CHSH inequalities only compare correlations
that are each between a term measured or potentially measured by Alice
and a term  measured or potentially measured by Bob.  This makes
(\ref{eq:CHSH}) important for experimental tests on Bell's theory.
But such experimental tests of (\ref{eq:CHSH}) are only meaningful
when assuming Locality (as will be detailed in \cite{dFT}). This gives
one full freedom  
in using whichever versions of (\ref{CHSH0}) are more suitable 
when dropping the Locality assumption.  
%%%%%%%%%%%%%%%%%%%%%%

Bell-type inequalities were discussed long ago by Boole,
\cite{Boole1854}, \cite{Boole1862} in a classical context (see
\cite{Pitowsky2001} and references therein).  The merit of Bell, who
did not know about Boole's work, is in part to have realized that such
inequalities arise when assuming some form of CMR to extend the usual
realm of QM: even if some of the mathematics of Boole inequalities was
known, the physics of the Bell inequalities and their violation under
QM, CMR and Locality assumptions are Bell's.  This being said, we shall
refer to the two inequalities written in compact form:  
\begin{equation}\label{Boole1}
|\inner{a}{b} \pm \inner{a}{a'}| + |\inner{b'}{b} 
\mp \inner{b'}{a'} |\leq 2\,. 
\end{equation}
as the \emph{Boole inequalities}. 

Locality allows one to compute all the pairwise correlations out of
the quadruplet $Q\equiv({\bm a}, {\bm b},{\bm a'}, {\bm b'})$: when 
assuming QM, CMR, and Locality, one can compute all correlations
$\inner{u}{v}$ where $u$ and $v$ stand for any of $a$, $a'$, $b$
and $b'$. Bell used that to compute:  
\begin{equation}\label{Correl3}
\inner{a}{a'} = {\bm a}\cdot {\bm a'} = \cos (\theta _{\bm a}- \theta _{\bm a'})\,.
\end{equation}
in his 1964 seminal paper 
\cite{Bell}.  Using (\ref{Correl3}), he got
a counterexample to Equation (\ref{Bell0}), thus proving the first
version of the following result.  
\smallskip

\noindent
{\bf Bell's Theorem \cite{Bell}.} \emph{The conjunction of Classical Microscopic
  Realism and Locality is incompatible with Quantum Mechanics}.  
\smallskip

At the end of \cite{Bell} Bell pointed out that the contradiction that he
had obtained would be compromised by abandoning the Locality hypothesis.
Indeed the main tool to go beyond
(\ref{Correl1}) and (\ref{Correl2}) simply disappears when Locality is
no longer assumed.  Besides making the evaluation of some correlations difficult,
abandoning Locality has a profound impact: sequences such as $\{a_i\}$
and $\{b_i\}$, as well as sequences such as $\{a'_i\}$ and $\{b'_i\}$
when assuming CMR, do not make sense anymore, contrary to an
\emph{a-priori} on so-called \emph{elements of Reality}  expressed in
\emph{EPR}.  Instead we need to consider objects that embody some
non-locality, such as the sequences $\{a_i ({\bm a}, {\bm b})\}$, $\{
b_i ({\bm a}, {\bm b})\}$, $\{a'_i ({\bm a'}, {\bm b'})\}$, and
$\{b'_i ({\bm a'}, {\bm b'})\}$.  We can still use QM to get
(\ref{Correl1}) and the conjunction of CMR and QM  to get
(\ref{Correl2}) without assuming Locality nor anything to replace that
hypothesis.  But in order to get a violation in the version of
(\ref{CHSH0}) that one uses, one also needs to know something about
all the correlations in that version.  
In particular, in order to extract a contradiction out of (\ref{Boole1}), 
we will need to gather enough information about the two correlations 
${\mathcal A}(Q)\equiv \inner{a ({\bm a}, {\bm b})}{ a' ({\bm a'}, {\bm b'})}$ 
and ${\mathcal B}(Q)\equiv\inner{b({\bm a}, {\bm b})}{ b' ({\bm a'}, {\bm b'})}$ 
(in short $\inner{a}{a'}$ and $\inner{b}{b'}$).  
  
We introduce here a two-parameter family ${\mathcal F}_{a,b}$ of quadruplets that
satisfy  $\theta _{\bm a'}+ \theta _{\bm a}=\theta_{\bm b'}+ \theta
_{\bm b}=0$, and in there, focus on special points $Q$ and on a
perturbation of the points $(\frac{\pi}{4},-\frac{3\pi}{4})$ and
$(\frac{3\pi}{4},-\frac{\pi}{4})$ along a path out of one-parameter
sub-family ${\mathcal F'}_{a}\subset {\mathcal F}_{a,b}$ where
furthermore  
$\theta_{\bm a}-\theta_{\bm b}=\pi$.  Figure 1 displays values of the
pairs $({\mathcal A}(Q), {\mathcal B}(Q))$ for a discrete set of
quadruplets of vectors determined by the pairs $({\bm a}, {\bm b})$
that unequivocally represents $Q$ in the family ${\mathcal  F}_{a,b}$
across a quarter of that family (by abuse of notation, we will also
call $Q$ the {\it pair of angles\/} that represents $Q$, \emph{e.g.,}
in Figure 1 and more generally in ${\mathcal  F}_{a,b}$).  There is no
need to extend ${\mathcal  F}_{a,b}$ beyond $[0,\pi]\times [-\pi,0]$
since the rest of the family can as well be obtained by exchanging
$a\leftrightarrow a'$ and/or $b\leftrightarrow b'$. 
%%%%%%%%%%%%%%%%%%START FIGURE%%%%%%%%%%%%%%%
\begin{figure}[h]
\centerline {\includegraphics[width=14cm]{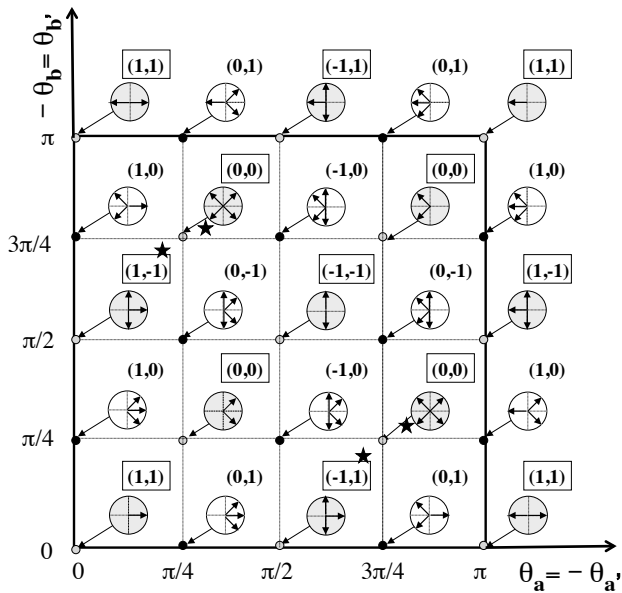}} 
\caption{The family ${\mathcal F}_{a,b}$ for $(\theta_{\bf
    a},-\theta_{\bm b})\in [0,\pi]^2$. The pairs are the values of
  $\inner{a}{a'}$ and $\inner{b}{b'}$.  They correspond to
  exact computations if they are framed, and to reasonable expectation
  of the written values otherwise. More symbols are discussed in the
  main text.} \label{Figur}  
\end{figure}
%%%%%%%%%%%%%%%%%%END FIGURE%%%%%%%%%%%%%%%

The framed pairs $({\mathcal A}(Q), {\mathcal B}(Q))$ with entries in
$\{-1,0,1\}$ correspond to exact and often rather easy computations if
no entry is zero: one then uses that ${\bm a'}={\bm a}$ and ${\bm
  b'}={\bm b}$ yields $(1,1)$ and that replacing ${\bm c}\in \{{\bm
  a}, {\bm b}\}$ by ${\bm{-c}}$ is not a change of instrument setting
but only a change of the signs of all readings along ${\bm c}$. The
points
$Q_{\frac{\pi}{4},-\frac{3\pi}{4}}=(\frac{\pi}{4},-\frac{3\pi}{4})$
and
$Q_{\frac{3\pi}{4},-\frac{\pi}{4}}=(\frac{3\pi}{4},-\frac{\pi}{4})$ are 
treated using the parity argument from \cite{Tresser1} and \cite{Tresser2}: 
since the configurations at both Alice's and Bob's sites are parity invariant, 
the corresponding pairs $({\mathcal A}(Q), {\mathcal B}(Q))$ are 
proved to be $(0,0)$ (Claims 3 and 4 below do not depend on any $Q$ with 
$({\mathcal  A}(Q), {\mathcal B}(Q))=(0,0)$).  The rigorously established pairs 
(white points in Figure 1) are consecutive along the \emph{anti-diagonal direction} 
$\{ \theta_{\bm a}-\theta_{\bm b}=\mathrm{const.} \}$ and along the 
\emph{main diagonal direction} $\{ \theta_{\bm a}+\theta_{\bm b}=\mathrm{const.} \}$.  
Some but not all the points where we have exact values are possibly 
compromise if contextuality holds true.
  
For the other pairs (black points), the values $\pm 1$ seem reasonable
by contemplating exact values on the same line $\theta_{\bm a}$ or
$\theta_{\bm b}$ equal to a constant while the values zero are
obtained by the parity argument mentioned above for
$Q_{\frac{\pi}{4},-\frac{3\pi}{4}}$ and
$Q_{\frac{3\pi}{4},-\frac{\pi}{4}}$.  However, the validity of the
parity-based argument now does depend on the (weak) assumption that the asymmetric
settings of Alice and Bob cannot break the parity symmetry used for the other observer.  
\begin{claim} If values in unframed pairs $({\mathcal A}(Q), {\mathcal B}(Q))$ 
are correct, then one gets a violation of a Boole inequality
that reads $\sqrt{2}+1\leq 2$, \emph{i.e.,} the error on  $|{\mathcal A}(Q)|+|{\mathcal B}(Q)|$ 
would need to be over $40\%$ to avoid the
violation of (\ref{Boole1}) if the pairs $({\mathcal A}(Q), {\mathcal
  B}(Q))$ are only approximately correct. 
\end{claim}
This follows from (\ref{Correl1}), (\ref{Correl2}), and the data in
Figure 1.  As for these data, notice that the  
unframed pairs can also be obtained as the arithmetic means of
the exact values at pairs of points that are on both sides of the
point under consideration, either on a vertical or on a horizontal line.  
%%%%%%%%%%%%%%%%
This is compatible with (\ref{CorrelLinear}) that was already discussed
as equation (10) in \cite{Bell}:
\begin{equation}\label{CorrelLinear}
\inner{a}{a'} =
1-\frac{2|\theta_a-\theta_{a'}|}{\pi}\,\,;\,\,\inner{b}{b'} =
1-\frac{2|\theta_b-\theta_{b'}|}{\pi}\,. 
\end{equation}
But setting $a=a'$ or $b=b'$ in (\ref{CorrelLinear}) implies Locality
in view of Claim \ref{Miracle} below.

\medskip
Let us now formulate two weak hypotheses that seem quite reasonable:
\begin{description}
\item[H1=Regularity.] ${\mathcal A}$ and ${\mathcal B}$ are real analytic functions of $\theta_{\bm a}$ and $\theta_{\bm b}$ off the pre-images of the extremal values.  
\item [H2=Transversality.] 
Away from points where both $\mathcal{A}$ and $\mathcal{B}$ are extremal, the critical values of $\mathcal{A}- \mathcal{B}$ in its domain of analyticity are bounded away from 0.
\end{description} 
Both {\bf H1} and {\bf H2} would appear as natural assumptions to most physicists.  We notice that instead of the real analyticity stated in {\bf H1}, $C^2$ smoothness off the pre-images of the extremal values would suffice 
to prove Claim 2. Also recall that $\inner{a}{b}$ and $\inner{a'}{b'}$ are analytic functions given by (\ref{Correl1})
and (\ref{Correl2}).  
 
Let us now perturb now away from $Q_{\frac{\pi}{4},-\frac{3\pi}{4}}$ and $Q_{\frac{3\pi}{4},-\frac{\pi}{4}}$ where $({\mathcal A}(Q),{\mathcal B}(Q))=(0,0)$ to
$Q_{\frac{\pi}{4}+\mu\epsilon,-\frac{3\pi}{4}+\nu\epsilon}$ and $Q_{\frac{3\pi}{4}+\mu\epsilon,-\frac{\pi}{4}+\nu\epsilon}$ with $\epsilon>0$ and $\mu,\nu \neq 0$ (the stars in Figure 1 locate $(\mu,\nu)= (\pm 2,\mp 1)$ for some $\epsilon>0$).  Using {\bf H1} and {\bf H2}, we get that ${\mathcal  A}$ and ${\mathcal  B}$ have a non-trivial first-order dependency on $\epsilon$.  Hence Claim \ref{lem2} below whose proof will use the following \emph{Simple Fact}:

\begin{lemma}
Let $\alpha,\beta$ be real numbers with $|\alpha|<1$, $|\beta|<1$ and
$\alpha\neq \beta$.  Then exactly one of the two inequalities $|1\pm
\alpha|+|1\mp \beta| \leq 2$ is false. %%%\qed 
\end{lemma}

\begin{claim}\label{lem2} Assume that {\bf H1} and {\bf H2} hold true
  near $Q_{x_0,y_0}\in
  \{Q_{\frac{\pi}{4},-\frac{3\pi}{4}},Q_{\frac{3\pi}{4},-\frac{\pi}{4}}\}\subset
     {\mathcal  F}_{a,b}$.  Then for a perturbation $(\mu\cdot\epsilon, \nu\cdot\epsilon)$ with
$\mu,\nu\neq 0$ of said $Q_{x_0,y_0}$ to
$Q_{x_0+\mu\epsilon,\,y_0+\nu\epsilon}$ for $\epsilon > 0$ small
enough, one of the inequalities (\ref{Boole1}) leads to a
contradiction. It follows that at least one of QM, CMR, {\bf H1} and {\bf H2} must fail to hold true. 
\end{claim}
\noindent
The proof of Claim 2 goes as follows.  At $Q_{x_0+\mu\epsilon,y_0+\nu\epsilon}$, from (\ref{Correl1}) and (\ref{Correl2}), we have:
\begin{equation}
\inner{a}{b} = \inner{a'}{b'} =1-\frac{{(\mu+\nu)^2\cdot
    \epsilon}^2}{2} + O(\epsilon^3)\,. 
\end{equation}
Using {\bf H1} and  {\bf H2} we also have:
\begin{equation}\label{Taylor}
\left\{ \begin{array}{lcl}  \inner{a}{a'} & = & \mathcal{A}(x_0+\mu\epsilon,y_0+\nu\epsilon) = 
(\alpha_1\mu+\alpha_2\nu)\epsilon + O(\epsilon^2), \\
                          \inner{b}{b'}
                        &  = & \mathcal{B}(x_0+\mu\epsilon,y_0+\nu\epsilon) =  
(\beta_1\mu+\beta_2\nu)\epsilon + O(\epsilon^2) \end{array}
\right.
\end{equation}
where $\alpha_1,\alpha_2,\beta_1,\beta_2$ are constants. There must exist at least
one pair $(\mu,\nu)$ for which $\alpha_1\mu+\alpha_2\nu \neq
\beta_1\mu+\beta_2\nu$. Otherwise, we would have
$\alpha_1-\beta_1=\alpha_2-\beta_2=0$ which would imply that
$\mathcal{A}-\mathcal{B}$ has a critical point at $(x_0,y_0)$ with
critical value $0$, contrary to {\bf H2}.  For this choice of
$(\mu,\nu)$ and $\epsilon$ small enough, 
$\alpha=\inner{a}{a'}$ and $\beta=\inner{b}{b'}$ satisfy the hypotheses
of the Simple Fact (up to an inessential second-order correction in $\epsilon$). 
Therefore one of the inequalities in (\ref{Boole1})
must be violated.  Q.E.D.

\medskip
In fact, we have discovered a wider range of hypotheses that let one arrive at 
further contradictions out of as weak as possible complements to [QM]+[CMR], the conjunction of QM and CMR. 
All such hypotheses (some needing too many words to fit in this letter) are either {\it weaker than\/} or {\it quite different from\/} the usual Locality 
hypothesis, or {\it both\/}. But they still allow us to prove that one of Boole's inequalities 
is violated. 

Here is one more flavour: We can study the correlations $\inner{a}{a'}$ and $\inner{b}{b'}$ in the space of {\it all\/} $Q$'s, in a neighborhood 
of the point $Q^0$ defined by $\theta_{\bm a}=\theta_{\bm a'}=\theta_{\bm b}=\theta_{\bm b'}=0$, where $\inner{a}{a'}=1=\inner{b}{b'}$. 
We regard both correlations as non-constant analytic functions of the 6 expressions
$x_1=|\theta_{\bm a}-\theta_{\bm a'}|,\,x_2=|\theta_{\bm b}-\theta_{\bm b'}|,\dots ,\,x_6=|\theta_{\bm a'}-\theta_{\bm b}|$, 
an hypothesis we call {\bf H1'}.  Using {\bf H1'}
we write the Taylor series of 
$\inner{a}{a'}$ in a neighborhood of $Q^0$ as:
\begin{equation}\label{correlaa}
 \inner{a}{a'}= 1 + P(x_1,x_2,\dots,x_6) +  R(x_1,x_2,\dots,x_6) 
\end{equation}
where $P$ is a homogeneous polynomial of the lowest possible degree, say $n$, and the remainder $R$ contains all other monomials, 
of degree $\geq n+1$.  The homogeneous polynomial $P$ reads:
\begin{equation}\label{poly}
 P\;=\;\sum c_{j_1,j_2,j_3,j_4,j_5,j_6}\, x_1^{j_1}x_2^{j_2}x_3^{j_3}x_4^{j_4}x_5^{j_5}x_6^{j_6} \ ,
\end{equation}
where the sum is over all sets of non-negative indices such that $\sum_{i=1}^{6} j_i = n$. 
Note that the Alice$\leftrightarrow $Bob symmetry exchanges $x_1\leftrightarrow x_2$, $x_3\leftrightarrow x_4$, 
$x_5\leftrightarrow x_6$. Thus, the formula for correlation $\inner{b}{b'}$ is obtained from \eqref{correlaa} 
by performing such exchanges. 

We also assume that the coefficient $c_{n,0,0,0,0,0}$ of $x_1^n$ in \eqref{poly} dominates all the others, an 
hypothesis we call {\it Local Dominance\/} ({\bf LD}). Hypothesis {\bf LD} is one possible way to express the 
physically plausible assumption that: \emph{``the correlation $\inner{a}{a'}$ should have a more sensitive dependency on 
$x_1=|\theta_{\bm a}-\theta_{\bm a'}|$, the local angle for Alice, than on all the more distant angles
(and similarly for $\inner{b}{b'}$ and Bob)"}. The following result will be proved in~\cite{dFT}.

\begin{claim} From \emph{[QM]+[CMR]+[{\bf H1'}]+[{\bf LD}]}, 
it follows that there are points $Q$ near $Q^0$ where one of the Boole inequalities (\ref{Boole1}) is violated. 
Hence, at least one of QM, CMR, {\bf H1'} and {\bf LD} must fail to hold true. 
\end{claim} 
The proof of Claim \ref{Miracle} below is straightforward.  
Yet the form of Locality met in Claim \ref{Miracle} is sufficient to develop Bell's theory. 
\begin{claim}\label{Miracle} Assume a theory {\bf T} that duplicates all the predictions of QM but satisfies CMR 
(\emph{i.e.,} the sequences $\{a_i\}$, $\{a'_i\}$, $\{b_i\}$ and $\{b'_i\}$ are well defined at once).  
If in {\bf T}, $\inner{a}{a'}$ and $\inner{b}{b'}$ are such that for any $c$ in $\{a,b\}$,
${\bf c}={\bf c'} \Rightarrow \inner{c}{c'}=1$, then the theory {\bf T}
is \emph{Local} in the sense that the sequence $\{a_i\}$ (resp. $\{b_i\}$) is independent of the choices of vectors made by $Bob$ (resp. Alice).
\end{claim}

The work reported here was initiated as follows.  Further progress made upon \cite{Tresser1} and 
\cite{Tresser2} revealed that the hypothesis that we were using, beside QM and CMR, was mostly a 
consequence of Special Relativity, leading us to hope that perhaps no extra hypothesis is in fact 
needed to reach a Bell inequality violation. Yet a classical argument against the sufficiency of [QM]+[CMR] 
to get a Bell Theorem is the existence of \emph{Bohmian Mechanics} (\emph{BM}), a PHV theory
that duplicates all the predictions of QM  \cite{Bohmian}.  We noticed however that 
\emph{the realist aspects of BM have not been as thoroughly investigated as the 
comparison of BM with QM where QM has a say.}  We thank Shelly Goldstein for explaining to us why 
%$\inner{a}{a'}=1-\frac{2|\theta_{\bf a}-\theta_{\bf a'}|}{\pi}$ 
$\inner{a}{a'}=1-{2|\theta_{\bf a}-\theta_{\bf a'}|}/{\pi}$ 
in BM, assuming that for each $i$ 
Alice makes her $i^{\rm th}$ measurement before Bob does his one in the Laboratory frame. 
Our findings do not permit such a structure of the correlation ${\mathcal A}(Q)$, at least  if we admit that the space time is actually relativistic, a viewpoint that strict Bohmians might not accept here;  
for now we will let readers draw their own conclusions.

In summary, while we have not yet succeeded in proving [QM]+[CMR] $\Rightarrow$ \emph{Contradiction}, 
we have shown here that there is a new collection of small sets of very weak hypotheses, anyone of 
these sets being called $``{\bf Some\,\, H}"$ for now, many of which are trivially 
weaker than Locality. Yet any such {\bf Some H} suffices to suggest and often let one prove: 
%\begin{equation}
$
{\rm [QM]+[CMR]}+{\bf Some\,\, H}\,\,\Rightarrow\,\, \emph{Contradiction}\,.
$
%\end{equation}
All that will be expanded, detailed and applied in \cite{dFT}.  

\medskip
%%%%%%%%%%%%%%%%%%%%%
\begin{acknowledgments}
\textbf{Acknowledgments:} 
This work was made possible by support from FAPESP through ``Projeto Tem\'atico {\it Din\^amica 
em Baixas Dimens\~oes\/}'', Proc. FAPESP 2006/03829-2, and also Proc. FAPESP 2011/50128-8.  Communications with S. Goldstein, R. Griffiths, P. Hohenberg, L. Horwitz, and E. Spiegel are gratefully acknowledged, and are hopefully the seeds of further work.
\end{acknowledgments}
%%%%%%%%%%%%%%%%%%%%%%%%%%%%%%%

\end{document}